\documentclass[12pt]{article}      
\usepackage{graphicx}

\usepackage{amsmath,amsfonts,amssymb}
\newtheorem{theorem}{Theorem}
\newtheorem{definition}{Definition}
\newtheorem{proposition}{Proposition}
\newtheorem{corollary}{Corollary}
\newtheorem{lemma}{Lemma}

\begin{document}

\title{Evolution of cooperation in a particular case of the infinitely repeated Prisoner's Dilemma with three strategies
}


\author{Irene N\'u\~nez Rodr\'{\i}guez         \and
        Armando G. M. Neves 
}

\author
{Irene N\'u\~nez Rodr\'{\i}guez$^{1}$ \\ Armando G. M. Neves$^{1}$
	\\
	\normalsize{$^{1}$Departamento de Matem\'atica, Universidade Federal de Minas Gerais,}\\
	\normalsize{irene.biobio@gmail.com}\\
	\normalsize{aneves@mat.ufmg.br}\\
	\normalsize{Av. Ant\^onio Carlos 6627, 30123-970 Belo Horizonte - MG, Brazil}
}

\maketitle

\begin{abstract}We will study a population of individuals playing the infinitely repeated Prisoner's Dilemma under replicator dynamics. The population consists of three kinds of individuals using the following reactive strategies: ALLD (individuals which always defect), ATFT (almost tit-for-tat: individuals which almost always repeat the opponent's last move) and G (generous individuals, which always cooperate when the opponent cooperated in the last move and have a positive probability $q$ of cooperating when they are defected). Our aim is studying in a mathematically rigorous fashion the dynamics of a simplified version for the computer experiment in \cite{NowSigNature} involving 100 reactive strategies. We will see that as the generosity degree of the G individuals varies, equilibria (rest points) of the dynamics appear or disappear, and the dynamics changes accordingly. Not only we will prove that the results of the experiment are true in our simplified version, but we will have complete control on the existence or non-existence of the equilbria for the dynamics for all possible values of the parameters, given that ATFT individuals are close enough to TFT. For most values of the parameters the dynamics will be completely determined.
	
\end{abstract}

\section{Introduction}
\label{intro}

We know that cooperation either between individuals, or between parts, exists not only in human societies, but, more in general, in all biological systems. Cooperating individuals usually have to pay a cost for the benefit of other individuals. It is then an interesting question to understand how cooperation can evolve in the light of Darwinian natural selection. Sigmund, Nowak and collaborators have studied in several contributions the evolution of cooperation, see e.g. \cite{Nowbook} and \cite{Now1} for some basic information and more references.

The essence of the problem can be grasped by the famous Prisoner's Dilemma (PD) stated in different forms by \cite{whywehelp} and \cite{calcselfishness}. In mathematical terms, the PD is characterized by only two strategies, a pay-off matrix
\begin{equation}  \label{simplepd}
A \,=\, \left( \begin{array}{cc}
R&S\\
T&P  \end{array} \right)\;,
\end{equation} 
and the pay-off rankings $T>R>P>S$. Strategy 1 is C (cooperate) and strategy 2 is D (defect); matrix element $a_{ij}$ is the pay-off received by an individual playing strategy $i$ confronted by an individual playing strategy $j$, $i,j \in \{1,2\}$. With the above ranking, given the opponent's strategy, defection always has a larger pay-off than cooperation. Using the jargon of game theory \cite{Nowbook}, in the PD strategy D is a strict Nash equilibrium, whereas C is not a Nash equilibrium. Rational individuals must choose D in the PD.

A window for cooperation may be opened when individuals can perceive that pay-off $R$ for mutual cooperation is better than $P$ for mutual defection. This cannot happen for the simple PD above, because individuals will interact only once. If individuals are given the opportunity of interacting many times before they receive their pay-offs, then a reciprocity mechanism can favor cooperation. But in this case C and D are not the only possible strategies and we have the problem of selecting among a huge number of strategies combining C and D.

In the 1970's \cite{Axelrod} studied strategies for the repeated PD. He organized two tournaments of the repeated PD and in both realizations of the tournament the winner strategy was the simplest among all submitted strategies: \textit{tit-for-tat} (TFT). TFT is the strategy which repeats the previous movement of its opponent.

In Evolutionary Game Theory, see \cite{MaynardSmithPrice} and \cite{MaynardSmith}, pay-offs are viewed as biological fitness and strategies in a population having larger pay-offs have the tendency to increase their frequencies. Suppose we have a population of individuals playing $n$ strategies $s_1, \dots, s_n$ and let $x_i(t)$ be the fraction of the population occupied at time $t$ by individuals playing strategy $s_i$. Let also $\vec{x}$ denote the vector $(x_1, \dots, x_n)$. We define the fitness of strategy $s_i$ as 
\begin{equation}
\label{fitnessfi}
f_i(\vec{x})= (A \vec{x})_i\;,
\end{equation}
the $i$-th element of a matrix product, where now $A$ is an $n \times n$ pay-off matrix. The mean fitness of the population is then
\begin{equation}  \label{meanfitness}
\phi \,=\, \sum_{i=1}^n f_i(\vec{x}) x_i\;.
\end{equation}
Finally, population dynamics is naturally given by a system of ordinary differential equations (ODEs) of the form
\begin{equation}
\label{repleq}
\dot{x}_i \,=\, (f_i(\vec{x})- \phi) x_i\;,
\end{equation}
$i=1, 2, \dots, n$, which were introduced in \cite{TayJon} and are called \textit{replicator dynamics} equations. It can be shown \cite{HofSig} that the simplex \[S_n=\{\vec{x} \in \mathbb{R}^n\;; x_i \geq 0, \sum_{i=1}^n x_i=1\} \]
is invariant under the replicator dynamics. 

From now on we will consider the infinitely repeated Prisoner's Dilemma (IRPD) i.e. the limiting case $w=1$ in the repeated PD, and as admissible only \textit{reactive strategies} defined below: 
\begin{definition} \label{defreactive}
	An individual adopting reactive strategy $r(p,q)$ will choose between C or D at some round based only on his opponent's choice at the round before, according to the following stochastic rule: choose C with probability $p$ if the opponent chose C the round before; choose C with probability $q$ if the opponent chose D the round before. 	
\end{definition}
We may think of $p \in [0,1]$ as a \textit{loyalty} parameter and of $q \in [0,1]$ as a \textit{forgiveness} parameter. Some simple strategies are recognized as reactive:  ALLD, individuals which always defect, is denoted as $r(0,0)$ and TFT is $r(1,0)$.

Reactive strategies were defined in \cite{gamedynasp} and further studied in \cite{nowaktpb} and \cite{clevel}. It can be shown \cite{clevel} that if 
\begin{equation}  \label{markovconvcondition}
|(p-q)(p'-q')|<1
\end{equation}
then the mean pay-off per round for each player may be defined in terms of the equilibrium distribution of a Markov chain. This mean pay-off per round defines the \textit{deterministic} IRPD pay-off $E(s,s')$ for a player with reactive strategy $s \equiv r(p,q)$ against a player with reactive strategy $s'\equiv r(p',q')$:
\begin{equation}  \label{irpdpayoff}
E(s,s') \,=\, G_1 c c'+ (S-P) c + (T-P) c'+ P \;,
	\end{equation}
where 
\begin{equation}  \label{clevels}
c\,=\,\frac{(p-q) q'+q}{1-(p-q)(p'-q')}\;\;\;\mathrm{and}\;\;\;
c'\,=\,\frac{(p'-q') q+q'}{1-(p-q)(p'-q')} 
\end{equation}
are respectively the equilibrium probabilities that $s$ cooperates with $s'$ and vice-versa, and 
	\begin{equation} 
	\label{G1}
	G_1\,=\,(R-T)+(P-S)
	\end{equation}
is a parameter which will have great importance in this work.

Notice that, apart other unimportant cases, condition (\ref{markovconvcondition}) is not satisfied in the case where $s=s'=r(1,0)$, i.e. both players are TFT. This is due to the fact that the outcome of the game between two TFT individuals must depend not only on their loyalty and forgiveness, but also on their initial moves: they will remain forever in the CC state when both play C at the first move, or in state DD when both play D, or alternate between states CD and DC in the remaining cases. As a consequence, pay-off $R$ obtained by a perfect TFT against another perfect TFT if both start cooperating may become as low as $P$ if arbitrarily small amounts of ``noise'' are present.

Inspiration for this work was provided by the seminal computer experiment performed in \cite{NowSigNature}. In that paper authors took 99 reactive strategies randomly chosen in the square $[0,1] \times [0,1]$. To that sample, suggested by Axelrod's results, they added by hand strategy $r(0.99,0.01)$, an almost TFT (ATFT) strategy. TFT was not selected due to the above mentioned impossibility of defining its pay-off. All 100 strategies were considered as having equal fractions at the initial time and then evolution given by replicator dynamics (\ref{repleq}) was numerically evaluated. Results were illustrated at Fig. 1 in \cite{NowSigNature} and are described as follows:
\begin{enumerate}
\item Initially, strategies far from $r(0,0)$ have their frequencies strongly depleted and it seems that the strategy closest to ALLD will extinguish all the others.
\item After some time the frequency of ATFT starts increasing, and it looks like it will win the game.
\item After a lot more time the ATFT frequency decreases and a surprising strategy named \textit{generous TFT} (GTFT) finally drives all other strategies to extinction. In the experiment of \cite{NowSigNature}, performed with the same parameter values as Axelrod's tournaments, i.e. $T=5$, $R=3$, $P=1$, $S=0$, the winner was the strategy closest to $r(1, \frac{1}{3})$.
\end{enumerate}

GTFT seems to have been discovered in \cite{Molander} and rediscovered exactly in \cite{NowSigNature}. Molander defined GTFT as strategy $r(1,q)$ with $q$ \textit{close to}
\begin{equation}
\label{molandergtft}
q_{GTFT}= \min\{\frac{2R-S-T}{R-S}, \frac{R-P}{T-P}\} \;.
\end{equation}
GTFT is what we may call a genuine collaborative strategy, \textit{altruistic} indeed, as stated in \cite{Molander}. GTFT does more than TFT in reciprocating cooperation, it is also forgiving. 

Our objective in writing this paper is to give precise mathematical arguments supporting the results found in \cite{Molander} and in \cite{NowSigNature}. Molander's paper considers a situation in which there is no dynamics at all, only pairwise comparison between pay-offs obtained using different strategies in the IRPD. Moreover strategies considered in his paper are not reactive, but mixed strategies \cite{HofSig}. One important difference instead between this paper and \cite{NowSigNature} is that, unable at this time to prove results for numbers of strategies as large as 100, we simplify their model and consider the IRPD with only \textit{the three more prominent strategies} in their experiment. Moreover, our results will be valid for any suitable choice of the many parameters of the problem, not a fixed choice, and any initial conditions for the population.

More concretely, we will consider a population of individuals under the replicator dynamics (\ref{repleq}) playing the IRPD (pay-offs calculated by (\ref{irpdpayoff}) and (\ref{clevels})) with three reactive strategies:
\begin{itemize}
\item Strategy 1 is an arbitrary ATFT, i.e. strategy $r(1-\epsilon_1, \epsilon_2)$, where $\epsilon_1$ and $\epsilon_2$ are positive and small enough. Differently of \cite{NowSigNature}, we need not consider $\epsilon_1=\epsilon_2$.
\item Strategy 2 is ALLD, i.e. $r(0,0)$.
\item Strategy 3, which will be called \textit{Generous} (G), is $r(1,q)$, with $q>(\epsilon_1+\epsilon_2)^{1/2}$, i.e. perfectly loyal individuals and more forgiving than the considered ATFT. 
\end{itemize}
We will prove the existence of a maximum amount of forgiveness $q_{GTFT}$ and a region of initial conditions with positive area such that, as in \cite{NowSigNature}, only strategy 3 will survive after infinite time. But also we will see what happens for larger values of $q$ and find out that in some cases we may still have some weaker form of cooperation evolution.

We will now define what we mean by weaker forms of cooperation evolution, so that we may state our results. For the sake of the following definition, we will suppose that the initial condition to be used in the dynamics is random, taken with uniform probability among all possible triples $(x_1(0),x_2(0),x_3(0))$ in the simplex $S_3$.
\begin{definition}
	We will say that the population admits 
	\begin{itemize}
		\item \textit{full} evolution of cooperation if there is a positive probability that the dynamics will lead to extinction of all individuals adopting strategies ALLD and ATFT.
		\item \textit{partial} evolution of cooperation if there is a positive probability that the dynamics will lead to extinction of all individuals adopting ALLD, but ATFT and G will remain.
		\item \textit{weak} evolution of cooperation if there is a positive probability that the dynamics will lead to some condition where the G individuals are not extinct, but will coexist with ATFT and ALLD.
		\item \textit{no} evolution of cooperation if the dynamics leads to extinction of the G individuals with probability 1.
	\end{itemize}
\end{definition}

The methods used are exact calculations of the pay-off matrix and analysis of its entries. Some of the results depend on asymptotic analysis in parameters $\epsilon_1$ and $\epsilon_2$. For each $q$ we will determine all equilibria of the replicator dynamics in $S_3$, study the dynamics at the boundary of $S_3$ and classify the few compatible phase portraits in the interior of that region, using for this classification results in \cite{Zeeman} and \cite{Bomze83}. In most cases only a single phase portrait of the complete table in \cite{Bomze83} is compatible with the existent equilibria and dynamics at the boundary. In such cases the dynamics will be completely determined. In some other cases more than one phase portrait in \cite{Bomze83} will be found compatible, but although dynamics is not yet fully determined, we have some conjectures about it. 

The results are summarized in Tables \ref{tabG1small}, \ref{tabG1large} and \ref{tabG10} separated by the possible signs of $G_1$ defined by (\ref{G1}). In any case we will define thresholds $q_{red}$, $q_{green}$, $q_{blue}$ and $q_{black}$ and one equilibrium for the replicator dynamics with any of the above colors. For all possible intervals we will state which equilibria are present in the biological region $S_3$. The red equilibrium will be biological in all cases, as well as points $E_1$, $E_2$ and $E_3$. The tables also show which are the compatible phase portrait diagrams according to their numbering in \cite{Bomze83}, and the kind of resulting cooperation evolution, if known. A letter R following the number of a diagram means that the phase portrait is the one in the diagram, but with all orbits having the reverse orientation.

\begin{table}
	\caption{The bullets indicate which equilibria are biological at each interval in the case $G_1<0$. The red equilibrium and vertices $E_1$, $E_2$ and $E_3$ are biological at all intervals. Possible evolution of cooperation types are F (full), P (partial), W (weak) or N (no). A type followed by a ? means a conjectured result.}
	\label{tabG1small}      
	\begin{tabular}{cccccc}
		\hline\noalign{\smallskip}
		& Green & Blue & Black & Diagrams & Type  \\
		\noalign{\smallskip}\hline\noalign{\smallskip}
		$((\epsilon_1+\epsilon_2)^{1/2},q_{green})$& &\textbullet  & & 38R & F \\
		$(q_{green}, q_{blue})$ & \textbullet & \textbullet &  & 34R & P\\
		$(q_{blue},q_{black})$ & \textbullet & & & 36 & P\\
		$(q_{black},1]$ & \textbullet & & \textbullet &  12, 12R, 13 & W? \\
		\noalign{\smallskip}\hline
	\end{tabular}
\end{table}

\begin{table}
	\caption{The bullets indicate which equilibria are biological at each interval in the case $G_1>0$. The red equilibrium and vertices $E_1$, $E_2$ and $E_3$ are biological at all intervals. Possible evolution of cooperation types are F (full), P (partial), W (weak) or N (no). A type followed by a ? means a conjectured result.}
	\label{tabG1large}      
	\begin{tabular}{cccccc}
		\hline\noalign{\smallskip}
		& Green & Blue & Black & Diagrams & Type  \\
		\noalign{\smallskip}\hline\noalign{\smallskip}
		$((\epsilon_1+\epsilon_2)^{1/2},q_{black})$& &\textbullet  &  & 38R & F\\
		$(q_{black}, q_{blue})$ &  & \textbullet &  \textbullet & 9R & F\\
		$(q_{blue},q_{green})$ &  & & \textbullet &  15R & N \\
		$(q_{green},1]$ & \textbullet & & \textbullet &  12, 12R, 13 & N?\\
		\noalign{\smallskip}\hline
	\end{tabular}
\end{table}

\begin{table}
	\caption{The bullets indicate which equilibria are biological at each interval in the case $G_1=0$. The red equilibrium and vertices $E_1$, $E_2$ and $E_3$ are biological at all intervals. Possible evolution of cooperation types are F (full), P (partial), W (weak) or N (no). A type followed by a ? means a conjectured result.}
	\label{tabG10}      
	\begin{tabular}{cccccc}
		\hline\noalign{\smallskip}
		& Green & Blue & Black & Diagrams & Type  \\
		\noalign{\smallskip}\hline\noalign{\smallskip}
		$((\epsilon_1+\epsilon_2)^{1/2},q_{black})$& &\textbullet  &  & 38R & F\\
		$q_{black}$ &  &  &  & 45 & F \\
		$(q_{black},1]$ & \textbullet & & \textbullet &  12, 12R, 13 & W?\\
		\noalign{\smallskip}\hline
	\end{tabular}
\end{table}

The paper is organized as follows. In Section \ref{payoffsection} we introduce the pay-off matrix, define the biological region, the red, green and blue lines and all the equilibria, with some important notations. In Section \ref{entriespayoff} we prove some simpler properties of the entries of the pay-off matrix. Section \ref{secloceq} contains the most important results of this paper, where appearance or disappearance of the equilibria in the biological region are calculated according to the sign of $G_1$ and the interval to which the forgiveness $q$ of the G individuals belongs. Section \ref{secdyn} relates the biological equilibria with the phase portraits in \cite{Bomze83} and the corresponding types of evolution of cooperation. The paper is closed by a conclusions section.
\section{Pay-off matrix and notations}
\label{payoffsection}
Let $T$, $R$, $P$ and $S$ be the entries of the pay-off matrix for the simple PD, as in (\ref{simplepd}). By definition of the simple PD, these parameters obey inequalities
\begin{equation}   \label{trpsorder}
T>R>P>S \;.
\end{equation}
We will also assume the following inequalities as hypotheses for the results in this paper:
\begin{equation}
\label{supplineq}
P<\frac{S+T}{2}<R\;.
\end{equation} 
The upper bound for $(S+T)/2$ is a natural condition to ensure that alternating between C and D is not as good as a steady C for a pair of players, and has already appeared in \cite{Molander} and further works. The lower bound seems to be a novel condition necessary for some of the proofs. As a consequence of this novel assumption we will have in Proposition \ref{Frho} that any amount of forgiveness in a reactive strategy will result in a pay-off larger than $P$ for that strategy against itself.

The pay-off matrix for the IRPD among the three strategies ATFT, ALLD and G may be calculated in a lengthy but straightforward fashion using (\ref{irpdpayoff}) and (\ref{clevels}). If strategies are numbered as in Section \ref{intro}, the result is
\begin{equation}
\label{irpdpaymatrix}
A=\left( \begin{array}{ccc}
	F(\frac{\epsilon_1}{\epsilon_2}) & (1-\epsilon_2)P+\epsilon_2 S & a_{13}(q) \\
(1-\epsilon_2)P+\epsilon_2 T & P & (1-q)P+qT \\
a_{31}(q) & (1-q)P+qS & R
	\end{array} \right)\;,
\end{equation}
where
\begin{equation}
\label{a11}
F(\rho)=\frac{P\rho^2+(S+T)\rho +R}{(1+\rho)^2} \;,
\end{equation}
\begin{eqnarray}
\label{a13}
a_{13}(q)&=&G_1\frac{\epsilon_{1}^{2}(1-q)}{[q+(1-q)(\epsilon_1+\epsilon_2)]^2}\nonumber\\
&-& \frac{\epsilon_1}{q+(1-q)(\epsilon_1+\epsilon_2)}\frac{2R-S-T+(T-R)(\epsilon_1+\epsilon_2)}{1-\epsilon_1-\epsilon_2}\\
&+&\frac{R(1-\epsilon_2)-\epsilon_1 S}{1-\epsilon_1-\epsilon_2}\; \nonumber
\end{eqnarray}
and 
\begin{eqnarray}
\label{a31}
a_{31}(q)&=&G_1\frac{\frac{\epsilon_{1}^{2}}{1-\epsilon_1-\epsilon_2}}{[q+(1-q)(\epsilon_1+\epsilon_2)]^2}\nonumber\\
&-&\frac{\epsilon_1}{q+(1-q)(\epsilon_1+\epsilon_2)}\frac{2R-S-T-[\epsilon_1(T-P)+\epsilon_2(R-S)]}{1-\epsilon_1-\epsilon_2}\\
&+&\frac{R(1-\epsilon_2)-\epsilon_1 T}{1-\epsilon_1-\epsilon_2}\;. \nonumber
\end{eqnarray}
In (\ref{a13}) and (\ref{a31}), $G_1$ is the combination of parameters in (\ref{G1}).

It is known \cite{HofSig} that the simplex $S_n$ is invariant under replicator dynamics (\ref{repleq}) for any number $n$ of strategies. In our case, we will always consider $x_3=1-x_1-x_2$, and accordingly call \textit{biological region} the projection of the simplex $S_3$ onto the $(x_1,x_2)$ plane, i.e. the closed triangle $B$ with vertices $E_1 \equiv (1,0)$, $E_2\equiv (0,1)$  and $E_3 \equiv (0,0)$. The sides of $B$ will be denoted by $L_1$, $L_2$ and $L_3$, where $L_i$ is the side on which $x_i=0$. 

Let the fitnesses $f_i$ be defined by (\ref{fitnessfi}) and for $i \neq j$ denote
\[n_{ij}=\{(x_1,x_2)\in\mathbb{R}^2\;;\, f_i(x_1,x_2,1-x_1-x_2)=f_{j}(x_1,x_2,1-x_1-x_2)\}\;\]
the straight lines in which two fitnesses are equal. We will also denote $P_{ijk}$ the point at which the line $n_{ij}$ intercepts the $x_k=0$ line. Notice that using their definition above, coordinates for the $P_{ijk}$ can be easily calculated in terms of the entries in the pay-off matrix (\ref{irpdpaymatrix}).

From general arguments, see \cite{HofSig}, the equilibria for the replicator dynamics with three strategies can be:
\begin{itemize}
	\item Points in which only one strategy is present, i.e. the vertices $E_1$, $E_2$ and $E_3$ of $B$.
	\item Points in which one strategy is absent and the other two have the same fitness, i.e. $P_{123}$, $P_{132}$ and $P_{231}$, whenever they exist.
	\item One point in which all three strategies have the same fitness. This is the intersection of the three lines $n_{12}$, $n_{13}$ and $n_{23}$, whenever it exists, and will be denoted $Q$. Notice that if two among these lines cross at a point, then the third lines must also pass through this point.
\end{itemize}

Notice that replicator dynamics in everywhere well-defined in $\mathbb{R}^2$ and the above mentioned points are equilibria whether they lie in $B$ or not. The fact that we are only interested in the dynamics in $B$ motivates the following definition:
\begin{definition}  \label{defbio}
We will say that equilibria $P_{123}$, $P_{132}$ and $P_{231}$ are \textit{biological} whenever they lie in $B$, but not coincide with any of the vertices. We will say that equilibrium $Q$ is \textit{biological} whenever it lies in the interior of $B$.
\end{definition}

In the rest of this paper we will study whether each of the above mentioned equilibria is biological or not. These equilibria will be studied by locating the intersections of the $n_{ij}$ lines with the sides of $B$ and with each other. We found it simpler for the sake of the upcoming notations to give arbitrary color codes to each of the straight lines $n_{ij}$ and equilibria. Henceforth line $n_{12}$ will be referred to as the \textit{red line} and $P_{123}$, intersection of the red line with $x_3=0$, \textit{red equilibrium}. Similarly, $n_{13}$ and $P_{132}$ will be called respectively \textit{green} line and equilibrium and $n_{23}$ and $P_{231}$ will be called \textit{blue} line and equilibrium. The coexistence equilibrium $Q$ will be called \textit{black equilibrium}.

We will always be interested in positive values for $\epsilon_1$ and $\epsilon_2$, and most of the results will hold given that these parameters are small enough. We define then polar coordinates $r$ and $\theta$ in the $(\epsilon_1, \epsilon_2)$ plane, so that 
\begin{equation} \label{polarcoord}
 \epsilon_1= r \cos \theta \;\;\;\mathrm{and} \;\;\; \epsilon_2=r \sin \theta
\end{equation}
Throughout this paper, $r$ and $\theta$ will always be used with this meaning.

Many times we will use $\epsilon$ to refer to vector $(\epsilon_1, \epsilon_2)$. We define the phrase ``property P holds if $\epsilon$ is small enough" as meaning ``there exists $r_0>0$ such that property P holds if $0<r<r_0$".

The overwhelming majority of the intermediate and final results in this paper will hold if $\epsilon$ is small enough. From now on, as with (\ref{trpsorder}) and (\ref{supplineq}), we will assume as a hypothesis for the rest of this paper that $\epsilon$ is small enough. In the beginning we will be explicit in stating this hypothesis, because we want the reader to be aware of it, but with time we will be increasingly more relapse in reminding it.

In some instances we will also use the notation $O(r^{\alpha})$ standard in asymptotic analysis. For not letting any doubt about it, if $f$ is some function depending on $\epsilon$, we will write $f=O(r^{\alpha})$ if there exist $r_0>0$  and a constant $K$ independent of $r$ such that $|f/r^{\alpha}|<K$ for $0<r<r_0$.

\section{Some properties of the entries of the pay-off matrix}
\label{entriespayoff}

We start by considering the pay-off $F(\frac{\epsilon_1}{\epsilon_2})$ of strategy ATFT against itself, with function $F$ being given by (\ref{a11}).
\begin{proposition}\label{Frho}

\begin{itemize}
		\item[(i)] $F(0)=R$.
		\item[(ii)] $\lim_{x \rightarrow\infty}F(x)=P$.
		\item[(iii)] $F$ is a decreasing function in $[0,+\infty)$.
		\item[(iv)] There exist positive constants $K_1, K_2$ such that
		\begin{equation}
				\label{F-P}
				\frac{K_1}{r}\leq \frac{F(\frac{\epsilon_1}{\epsilon_2})-P}{\epsilon_2}\leq \frac{K_2}{r}\,.
				\end{equation}
				\item[(v)] There exist positive constants $K_3, K_4$ such that
				\begin{equation}
						\label{R-F}
						\frac{K_3}{r}\leq \frac{R-F(\frac{\epsilon_1}{\epsilon_2})}{\epsilon_1}\leq \frac{K_4}{r}\,.
						\end{equation}
	\end{itemize}
\end{proposition}
	
	\textbf{Proof}
The first two items are direct consequences of (\ref{a11}). The third item follows easily by calculating the derivative of $F$ and using both inequalities in (\ref{supplineq}).

Using polar coordinates (\ref{polarcoord}) we get
\begin{eqnarray*}
				\frac{F(\frac{\epsilon_1}{\epsilon_2})-P}{\epsilon_2}
				&=&\frac{1}{r} \; \frac{(R-P) \sin\theta+(S+T-2P)\cos\theta}{(\cos\theta+\sin\theta)^2}\;.
			\end{eqnarray*}
By using (\ref{trpsorder}) and (\ref{supplineq}) the function of $\theta$ in the right-hand side is clearly strictly positive and continuous in the compact $[0, \frac{\pi}{2}]$. Letting $K_1$ be its minimum and $K_2$ its maximum, assertion (iv) is proved.

Item (v) can be proved in an analogous way.  $\blacksquare$

Items (i), (ii) and (iii) in Proposition \ref{Frho} prove that the pay-off of an ATFT against another ATFT may be any number in $(P,R)$ regardless of the smallness of $\epsilon$. 

If $q=0$, strategy G becomes TFT. As a consequence of this, see \cite{clevel},
\begin{equation}  \label{a130}
a_{13}(0)= a_{31}(0) = F(\frac{\epsilon_1}{\epsilon_2}) \;.
\end{equation}

Also at $q=1$ both formulas for $a_{13}$ and $a_{31}$ simplify and we obtain
\begin{equation}   \label{a13a31q=1}
a_{13}(1)\,=\, R+(T-R)\epsilon_1 \;\;\;\mathrm{and}\;\;\;
a_{31}(1)\,=\,R-(R-S)\epsilon_1\;. 
\end{equation}

Other important properties of these same entries are:
\begin{proposition}\label{propa13a31}
	\begin{itemize}
		\item[(i)] $a_{13}(q)>a_{31}(q)\;\forall q\in (0,1]$.
		\item[(ii)] $a_{13}(q)-a_{31}(q)$ is an increasing function in $[0,1]$.
		\item[(iii)] If $\epsilon$ is small enough, then both $a_{13}(q)$ and $a_{31}(q)$ are increasing functions in $[0,1] $.
		\item[(iv)] $a'_{13}(0) \stackrel{r \rightarrow 0}{\rightarrow} \infty$ and $a'_{31}(0) \stackrel{r \rightarrow 0}{\rightarrow} \infty$.
		\item[(v)]  $a'_{13}(1) \stackrel{r \rightarrow 0}{\rightarrow} 0$ and $a'_{31}(1) \stackrel{r \rightarrow 0}{\rightarrow} 0$.
		\item[(vi)] If $\epsilon$ is small enough, $a''_{13}(q)$ and $a''_{31}(q)$ are both negative in $[0,1]$.
	\end{itemize}
\end{proposition}

\textbf{Proof}
After some easy manipulations with (\ref{a13}) and (\ref{a31}), we obtain
\[a_{13}(q)-a_{31}(q)=\frac{\epsilon_1(T-S)q}{q+(1-q)(\epsilon_1+\epsilon_2)}\;,\]
which proves assertion (i). Differentiating the above equation proves (ii).

To prove (iii), we define first an auxiliary variable
\begin{equation}
\label{defx}
x \equiv q + (1-q)(\epsilon_1+\epsilon_2)
\end{equation}
which leads us to
\begin{equation}  \label{da31}
a'_{31}(q) \,=\, \epsilon_1 \, \frac{2R-S-T- \epsilon_1(T-P) - \epsilon_2(R-S)}{x^2} \, -\, \frac{2G_1 \epsilon_1^2}{x^3} \;.
\end{equation}
Notice then that $a'_{31}(q)/ \epsilon_1$ is a continuous function of $\epsilon$, positive at $\epsilon=0$. As a consequence, $a'_{31}(q)$ is positive if $\epsilon_1>0$, provided that $\epsilon$ is small enough. Using (ii) the analog result is obtained for $a_{13}(q)$.

To prove (iv), we substitute $q=0$, thus $x= \epsilon_1+\epsilon_2$, in (\ref{da31}). Using polar coordinates and (\ref{G1}), we get
\[ a'_{31}(0) \,=\, \frac{1}{r} \, \frac{(S+T-2P) \cos^2 \theta+ (2R-S-T) \sin \theta}{(\cos \theta+ \sin \theta)^3} \, +\, O(1) \;.\]
As the function of $\theta$ multiplying $1/r$ is positive, then the result for $a'_{31}(0)$ is proved. Using again (ii), we prove the same for $a'_{13}(0)$.

The proofs of (v) and (vi) follow similar ideas.   $\blacksquare$

Although formulas (\ref{a13}) and (\ref{a31}) are complicated, Proposition \ref{propa13a31} tells a lot about these functions. In particular, properties (iv) and (v) show that both $a_{13}$ and $a_{31}$ grow very fast for $q$ close to 0 and then saturate before $q=1$.

To close this section, a simple and important 
\begin{corollary}
	\label{uniqueroot}
	If $\epsilon$ is small enough and $\alpha \in (F(\frac{\epsilon_1}{\epsilon_2}),R+(T-R)\epsilon_1)$, then equation  $a_{13}(q)=\alpha$ has a unique root $q$ in interval $(0,1)$.
	Analogously, if $\beta \in (F(\frac{\epsilon_1}{\epsilon_2}),R-(R-S)\epsilon_1)$, then $a_{31}(q)=\beta$ has a unique root in $(0,1)$.
	\end{corollary}
	
\section{Locating the equilibria}  \label{secloceq}
We start this section by studying the red equilibrium $P_{123}$, the simplest among the equilibria in which only two strategies coexist, because its location is independent of the variable $q$, as shown by the following result.
\begin{proposition}
	\label{P123}
	The red equilibrium $P_{123}$ is independent of $q$ and always biological.
	\end{proposition}
	\textbf{Proof}
Equating fitnesses $f_1$ and $f_2$, given by (\ref{fitnessfi}), and writing $x_2=1-x_1$, which is equivalent to $x_3=0$, we obtain
		\begin{equation}  \label{x1p123}
		x_1(P_{123})=\
	\frac{1}{1- \frac{T-P}{P-S}+\frac{1}{P-S} \, \frac{F(\frac{\epsilon_1}{\epsilon_2})-P}{\epsilon_2}} \;,
		\end{equation}
which is indeed independent of $q$. Using (iv) in Proposition \ref{Frho} we see that the denominator in the above equation is dominated by a positive term of order $1/r$. Thus $x_1(P_{123})>0$ and as small as we want if $\epsilon$ is small enough.   $\blacksquare$

The next result will be important when showing that the black equilibrium will become biological for some intervals in $q$, because it states that the intercepts of lines $n_{12}$, $n_{13}$ and $n_{23}$ appear always in the same order on $L_3$. Notice the appearance for the first time of a hypothesis stating that forgiveness $q$ of individuals adopting strategy 3 must not be too close to 0. This will happen in other parts of this section and has the clear meaning that strategy 3 must be more forgiving than strategy 1 for some of the results to be true.

\begin{theorem}[Order on $L_3$]
	\label{orderL3}
	If $q\in [(\epsilon_1+\epsilon_2)^{1/2},1]$, then $0< x_1(P_{123}) <x_1(P_{233}) < x_1(P_{133})<1$.
\end{theorem}
	\textbf{Proof}
For ease of comparison, we may rewrite all three quantities in the form $x_1(P_{ij3})= \frac{1}{1+\tilde{c}_{ij}}$, where formulas for the $\tilde{c}_{ij}$ will be presented. We will show that $0<\tilde{c}_{13}<\tilde{c}_{23}<\tilde{c}_{12}$, from which the claim will be a trivial consequence.

An easy calculation leads to
$$\tilde{c}_{13}=\frac{a_{31}(q)-F(\frac{\epsilon_1}{\epsilon_2})}{q(P-S)}$$
and		$$\tilde{c}_{23}=\frac{a_{31}(q)-P-\epsilon_2(T-P)}{q(P-S)}\,$$
and $\tilde{c}_{12}$ may be obtained in (\ref{x1p123}).
As $a_{31}(q)$ is increasing and $a_{31}(0)= F(\frac{\epsilon_1}{\epsilon_2})$, then $\tilde{c}_{13}>0$. If the ratio $\frac{\epsilon_1}{\epsilon_2}$ is fixed and $\epsilon$ is taken small enough, we obtain, using (ii) and (iii) in Proposition \ref{Frho}, that $F(\frac{\epsilon_1}{\epsilon_2})> P + \epsilon_2 (T-P)$, thus proving that $\tilde{c}_{13}<\tilde{c}_{23}$ for small enough $\epsilon$ and $q>0$.

Using (\ref{a13a31q=1}) and, again, the fact that $a_{31}(q)$ is increasing, we may see that, if $q>(\epsilon_1+\epsilon_2)^{1/2}$,
\[\tilde{c}_{23} < \frac{R-(R-S)\epsilon_1-P-\epsilon_2(T-P)}{(\epsilon_1+\epsilon_2)^{1/2}(P-S)} \,,
\]
which increases as $r^{-1/2}$ when $r \rightarrow 0$. On the other hand, by (\ref{x1p123}) and (iv) in Proposition \ref{Frho}, we see that $\tilde{c}_{12}$ increases as $r^{-1}$. We conclude that $\tilde{c}_{23}<\tilde{c}_{12}$ for small enough $\epsilon$.   $\blacksquare$

We may now define two numbers related to when the green and blue equilibria become biological:
\begin{definition}  \label{defqgb}
	According to Corollary \ref{uniqueroot}, 
	\[a_{13}(q)=R\]
has a unique root in $(0,1)$. Let $q_{green}$ be this root.

Let also
\begin{equation}  \label{qblue}
q_{blue}= \frac{R-P}{T-P} 
\end{equation}
be the unique root of equation $a_{23}=R$.
\end{definition}

With these definitions we prove an important general result:
\begin{theorem}  \label{biolgb}
Let $q \in (\epsilon_2,1]$. Then:
\begin{itemize}
\item The blue line intercepts $L_1$ if and only $q\leq q_{blue}$ and intercepts $L_2$ if and only if $q \in [q_{blue},1]$. In particular, the blue equilibrium is biological if and only if $q < q_{blue}$.
\item The green line intercepts $L_1$ if and only $q\leq q_{green}$ and intercepts $L_2$ if and only if $q \in [q_{green},1]$. In particular, the green equilibrium is biological if and only if $q > q_{green}$. 
\end{itemize}
\end{theorem}
\textbf{Proof}
After easy calculations we get 
\begin{equation}
\label{x2p231}
x_2(P_{231}) \,=\, \frac{1}{1+ \frac{q(P-S)}{R-P-q(T-P)}}\;,
\end{equation}
\begin{equation}
\label{x1p232}
x_1(P_{232}) \,=\, \frac{1}{1- \frac{a_{31}(q)-P - \epsilon_2 (T-P)}{R-P-q(T-P)}}\;,
\end{equation}
\begin{equation}
\label{x1p132}
x_1(P_{132}) \,=\, \frac{1}{1+ \frac{a_{31}(q)-F(\epsilon_1/\epsilon_2)}{a_{13}(q)-R}} \;,
\end{equation}
and
\begin{equation}
\label{x2p131}
x_2(P_{131}) \,=\, \frac{1}{1 -  \frac{(q-\epsilon_2)(P-S)}{a_{13}(q)-R}} \;,
\end{equation}
all in the form $1/(1+X)$, which will be in $(0,1)$ if and only if the corresponding $X$ is positive. 
In all four cases the proof that the necessary $X$ is positive if and only if the respective condition on $q$ is satisfied is trivial. In the cases related to the green line, we must use item (iii) in Proposition \ref{propa13a31}.   $\blacksquare$

Besides $q_{blue}$ and $q_{green}$ we will define a number $q_{red}$, which will signal the passage of the red line through the origin. In order to do that, let $\mu(q)$ be defined by
\begin{equation}
\label{defmu}\mu(q)=a_{13}(q)-P-q(T-P)\,.
\end{equation}
In terms of this new function we may easily obtain
\begin{equation}
x_2(P_{121}) \,=\, \frac{\mu(q)}{\mu(q)+ \epsilon_2 (P-S)}
\end{equation}
and 
\begin{equation}  \label{x1p122}
x_1(P_{122}) \,=\, \frac{\mu(q)}{\mu(q)- [F(\epsilon_1/\epsilon_2)-P- \epsilon_2 (T-P)]} \;,
\end{equation}
which show that the red line passes through the origin of the $(x_1, x_2)$ plane whenever $\mu$ has a zero. 

The following lemma proves existence and uniqueness of such a zero:
\begin{lemma}
	\label{mufunction} 
	Function $\mu$ defined by (\ref{defmu}) has a single critical point $\overline{q}$ and a single zero $q_{red}$ in $(0,1)$ such that $q_{red}>\overline{q}$. Furthermore, $\overline{q}$ is a maximum point, $\mu$ is positive in $(0,q_{red})$ and negative in $(q_{red},1]$.
\end{lemma}
	\textbf{Proof}
		As $\mu'(q)=a'_{13}(q)-(T-P)$, then (iv) and (v) in Proposition \ref{propa13a31} imply that $\mu'(0)>0$ and $\mu'(1)<0$ if $\epsilon$ is small enough. Then $\mu$ has at least a critical point $\overline{q} \in (0,1)$. Item (vi) in the same proposition proves uniqueness for $\overline{q}$ and that it must be a maximum point.
		
		As $\mu(0)=F(\frac{\epsilon_1}{\epsilon_2})-P>0$, then $\mu(\overline{q})>0$. And as $\mu(1)=R-T-(T-R)\epsilon_1<0$, then $\mu$ has a single zero in $(0,1)$ and this zero is located in $(\overline{q},1)$. The assertion on the signs of $\mu$ follows from the fact $\mu'(q_{red})<0$.   $\blacksquare$

It is now time to start displaying important results in which the sign of $G_1$ defined in (\ref{G1}) plays an important role. The first thing to notice is that formula (\ref{a13}) for $a_{13}(q)$ is notably simplified when $G_1=0$. Solving equation $a_{13}(q)=R$ is trivial and we get, for $G_1=0$,
\[q_{green}= \frac{2R-S-T}{R-S}\;.\] 
If we calculate the difference between this value and $q_{blue}$ we discover the identity
\begin{equation}
\label{difgreenblue}
\frac{2R-S-T}{R-S} - q_{blue} \,=\, \frac{G_1(T-R)}{(R-S)(T-P)}\;,
\end{equation}
which shows that $q_{green}$ and $q_{blue}$ coincide when $G_1=0$. As $a_{13}(q_{green})=R$, we discover that $\mu(q_{green})= (q_{blue}-q_{green})(T-P)$, from which we can deduce that $q_{red}$ also coincides with $q_{green}$ and $q_{blue}$ when $G_1=0$.

If $G_1 \neq 0$, although more complicated, equation $a_{13}(q)=R$ leads only to a second-degree equation in $q$ and a closed formula for $q_{green}$ can also be obtained. If we solve the equation in the auxiliary variable $x$ defined in (\ref{defx}) and notice that $q$ and $x$ differ by $O(r)$, we prove in general that
\begin{equation}
\label{explqgreen}
q_{green}=\frac{2R-S-T}{R-S}-\frac{G_1(T-R)}{(R-S)(2R-S-T)}\epsilon_1+O(r^2)\;,
\end{equation}
with the interesting consequence that the exact calculated value of $q_{green}$ for $G_1=0$ holds as a good approximation for $q_{green}$ even when $G_1 \neq 0$. 

By using the ideas above we can easily prove
\begin{theorem}
\label{g1andorder}
\begin{itemize}
	\item[(i)] If $G_1=0$, then $q_{green}=q_{blue}=q_{red}$.
	\item[(ii)] If $G_1<0$, then $q_{green}<q_{blue}<q_{red}$.
	\item[(iii)] If $G_1>0$, then $q_{green}>q_{blue}>q_{red}$.
\end{itemize}
\end{theorem}

Equation  (\ref{explqgreen}) will be useful later to guarantee that $q_{green}$ does not tend to 0 when $ r \rightarrow 0$. We will also need to prove the same for $q_{red}$. This is an easy consequence of the next result.
\begin{proposition}  \label{propestqred}
\begin{equation}
q_{red}=q_{blue}+O(r)  \label{estqred}
\end{equation}
\end{proposition}
\textbf{Proof}
	Using the definition of $\mu$ (\ref{defmu}) and (\ref{a13}), we may rewrite $\mu(q)=0$ as 
\begin{eqnarray*}
\frac{R(1-\epsilon_2)-\epsilon_1 S}{1-\epsilon_1-\epsilon_2}&-& \frac{\epsilon_1[2R-S-T+(T-R)(\epsilon_1+\epsilon_2)+G_1 \epsilon_1]}{1-\epsilon_1-\epsilon_2} \, \frac{1}{x}+ \frac{G_1 \epsilon_{1}^{2}}{1-\epsilon_1-\epsilon_2} \, \frac{1}{x^2}\\
	&=& P \,+\, \frac{x-\epsilon_1-\epsilon_2}{1-\epsilon_1-\epsilon_2}\,(T-P)\;, 
	\end{eqnarray*}
	which solution in $x$ will yield $q_{red}$. If we substitute $\epsilon_1=\epsilon_2=0$ above, the solution is simply $x_0=q_{blue}$.
	
	Substituting $x_0$ we may rewrite the above equation as $g(x,\epsilon)=0$, with
	\begin{eqnarray*}
	g(x, \epsilon)&=&  -\frac{\epsilon_1[2R-S-T+(T-R)(\epsilon_1+\epsilon_2)+G_1 \epsilon_1]}{1-\epsilon_1-\epsilon_2} \, \frac{1}{x}\\ &+& \frac{G_1 \epsilon_{1}^{2}}{x^2} \,+\, x_0-x + \epsilon_1(T-S)+ \epsilon_2(T-R) \;.
	\end{eqnarray*}
	As $g(x_0,0)=0$ with $\frac{\partial g}{\partial x}(x_0,0)=-1+O(r) \neq 0$ for small enough $r$, the implicit function theorem proves that in some neighborhood of $\epsilon=0$ the root $x$ of $g(x,\epsilon)=0$ is a differentiable function of $\epsilon$. Differentiability implies that this root is $x=x_0+O(r)$. Noticing that $q$ and $x$ differ by $O(r)$ leads to (\ref{estqred}).   $\blacksquare$

We will start justifying tables \ref{tabG1small} to \ref{tabG10} with the case $G_1>0$. Before that, a couple of technical results still independent of $G_1$.
\begin{lemma}
	\label{lemmaa31-F}
	Let $q_0 \in(0,1]$ be fixed and independent of $\epsilon$. Then:
		\begin{itemize}
		\item  $a_{31}(q_0)-F(\frac{\epsilon_1}{\epsilon_2})$ does not tend to 0 when $r \rightarrow 0$.
		\item If $q \in[q_0,1]$, $\frac{d}{dq}x_1(P_{132})= O(r)$.
	\end{itemize}
	 
	\end{lemma}
	\textbf{Proof}
		Using variable $x$, defined in (\ref{defx}), and (\ref{a31}) we may write
		\begin{eqnarray*}
			&&a_{31}(q)-F(\frac{\epsilon_1}{\epsilon_2}) \,=\, R-F(\frac{\epsilon_1}{\epsilon_2})+\frac{(R-T)\epsilon_1}{1-\epsilon_1-\epsilon_2}\\&-&\frac{\epsilon_1}{1-\epsilon_1-\epsilon_2}\frac{2R-S-T-[(T-P)\epsilon_1+(R-S)\epsilon_2]}{x}
			+\frac{G_1 \epsilon_{1}^{2}}{(1-\epsilon_1-\epsilon_2)x^2}\,.
		\end{eqnarray*}
		Remember that if $q_0$ is fixed, $x$ does not tend to 0 when $r \rightarrow 0$. This means that, apart of the term $R-F(\frac{\epsilon_1}{\epsilon_2})$, the remaining terms in the right-hand side do tend to 0 when $r \rightarrow 0$. On the other hand, (v) in Proposition \ref{Frho} proves that $R-F(\frac{\epsilon_1}{\epsilon_2})$ is positive and does not tend to 0. This proves the first part.
		
		In order to prove the second part, notice that
			\begin{eqnarray*}
				\frac{d}{dq}x_1(P_{132})&=&
				\frac{a'_{13}(q)\left(a_{31}(q)-F(\frac{\epsilon_1}{\epsilon_2})\right)+(R-a_{13}(q))a'_{31}(q)}{\left[a_{13}(q)-R+a_{31}(q)- F(\frac{\epsilon_1}{\epsilon_2})\right]^2}\,.
			\end{eqnarray*}
			If $q \geq q_{0}$, $a_{13}(q)-R$ is $O(r)$. Moreover, by the first part of this Lemma, $a_{31}(q)- F(\frac{\epsilon_1}{\epsilon_2})$ does not tend to 0 when $r \rightarrow 0$.  Thus the denominator in the expression above does not tend to 0 when $r \rightarrow 0$. In the numerator both terms are $O(r)$, as can be seen in (\ref{da31}) and its analog for $a_{13}$. As a consequence, the derivative of $x_1(P_{132})$ is $O(r)$.
		   $\blacksquare$

\begin{proposition}  \label{positiveder}
	If $q \in [q_{red},1]$, then 
	\[\frac{d}{dq}(x_1(P_{122})-x_1(P_{132}))>0\;.\]
\end{proposition}
\textbf{Proof}
	Formulas for $x_1(P_{122})$ and $x_1(P_{132})$ have already been given, see (\ref{x1p122}) and (\ref{x1p132}).
	
	Equation (\ref{estqred}) guarantees that $q_{red}$ does not tend to 0 as $r \rightarrow 0$. Then, by Lemma \ref{lemmaa31-F}, using e.g. $q_0=1/2 q_{blue}<q_{red}$, we conclude that $\frac{d}{dq}x_1(P_{132})$ is $O(r)$ in $[q_{red},1]$.
	
	By an easy calculation, we have
	\begin{equation}
	\frac{d}{dq}x_1(P_{122})
	\,=\,\frac{[(T-P)-a'_{13}(q)][F(\frac{\epsilon_1}{\epsilon_2})-P- \epsilon_2(T-P)]}{[F(\frac{\epsilon_1}{\epsilon_2})-P-\epsilon_2(T-P)-\mu(q)]^2}\;.
	\end{equation}
	It follows that
	\[
	\frac{d}{dq}x_1(P_{122}) \,>\, \frac{(T-P)-a'_{13}(q_{red})}{F(\frac{\epsilon_1}{\epsilon_2})-P-\epsilon_2(T-P)}\,,
	\]
	because in $[q_{red},1]$ we have $\mu(q)\leq0$, $a'_{13}(q)<a'_{13}(q_{red})$ and $F(\frac{\epsilon_1}{\epsilon_2})-P-\epsilon_2(T-P) >0$. In this last expression $a'_{13}(q_{red})$ is $O(r)$ and the denominator does not tend to 0. So there exists a positive constant $C$ independent of $r$ such that $\frac{d}{dq}x_1(P_{122})>C$.
	
	We conclude that for small enough $\epsilon$, $\frac{d}{dq}(x_1(P_{122})-x_1(P_{132}))>0$ for $q \in [q_{red},1]$.   $\blacksquare$

	We can now prove our first result on the black equilibrium:
\begin{proposition}  \label{posG1largeq}
	If $G_1 \geq 0$, the black equilibrium is biological for all $q \in (q_{red},1]$. If $G_1>0$ the conclusion extends also to $q=q_{red}$.
\end{proposition}
\textbf{Proof}
First of all, if $G_1 \geq 0$, by Theorem \ref{g1andorder} we know that $q_{red} \leq q_{green}$. For $q=q_{red}$ we then know that the green line intercepts the sides $L_3$ (Theorem \ref{orderL3}) and $L_2$ (Theorem \ref{biolgb}) and then we must have $x_1(P_{132})\leq 0$ for $q=q_{red}$. But $x_1(P_{122})=0$ at $q_{red}$. By Proposition \ref{positiveder} we discover that $x_1(P_{122})>x_1(P_{132})$ for $q \in(q_{red},1]$. Comparing this order with the order on side $L_3$ given by Theorem \ref{orderL3}, we see that the red and green lines must cross at the interior of $B$ for all $q \in(q_{red},1]$. If $G_1=0$ we already knew that the red and green lines crossed exactly at the origin for $q=q_{red}$. But for $G_1>0$, $x_1(P_{132})<0$ already at $q=q_{red}$, and the lines cross in the interior.  $\blacksquare$

Our task is now to find out when the black equilibrium first becomes biological and if it ever loses its biological status. We start with a general result:
\begin{proposition}  \label{ordsmallq}
	For any value of $G_1$ and $q=(\epsilon_1+\epsilon_2)^{1/2}$ we have $x_2(P_{121})>x_2(P_{131})$ .
\end{proposition}
\textbf{Proof}
The coordinates for the intercepts of the $n_{ij}$ lines with $x_1=0$ may all be written as
\begin{equation*}
x_2(P_{ij1})\,=\,\frac{1}{1+b_{ij}(q)}\,.
\end{equation*}
where, after easy calculations, we get
\begin{equation}  \label{x2pij1}
b_{12}(q)\,=\,\frac{\epsilon_2 (P-S)}{\mu(q)} \;\;\;\mathrm{and} \;\;\;
b_{13}(q)=\frac{(q-\epsilon_2)(P-S)}{R-a_{13}(q)}\,.
\end{equation}
As $a_{13}((\epsilon_1+\epsilon_2)^{1/2})= R- O(r^{1/2})$ and  $\mu((\epsilon_1+\epsilon_2)^{1/2})= R- P-O(r^{1/2})$, then by the above expressions our assertion is true.   $\blacksquare$

The order just proved between $x_2(P_{121})$ and $x_2(P_{131})$ at $q=(\epsilon_1+\epsilon_2)^{1/2}$ is reversed at $q=q_{red}$ if $G_1>0$. In fact, if $G_1>0$ and $q=q_{red}$, we have $x_2(P_{131})>0$, because $q_{green}>q_{red}$, and $x_2(P_{121})=0$. It turns out that the green and red lines must cross on $L_1$ for at least one $q \in ((\epsilon_1+\epsilon_2)^{1/2},q_{red})$. We will show that this crossing is indeed unique, defining $q_{black}$. If $G_1=0$, we saw that the green and red lines crossed at $q=q_{red}$. We will also show that no other crossing will happen if $q \in ((\epsilon_1+\epsilon_2)^{1/2},q_{red})$.

If $x_2(P_{121})-x_2(P_{131})$ were monotonic in $((\epsilon_1+\epsilon_2)^{1/2},q_{red})$ the assertions in the preceding paragraph would be trivial. As this does not happen, we must work a bit harder, starting with
\begin{proposition}
	\label{propqtilde}
	Equation
	\begin{equation}  \label{muqtil}
		\mu(q)= (\epsilon_1+\epsilon_2)^{1/2}
	\end{equation}
	has a single root $\tilde{q} \in [0,q_{red}]$, which is asymptotically given by
	\begin{equation}  \label{estqtil}
	\tilde{q} \,=\, q_{blue} - \frac{(\epsilon_1+\epsilon_2)^{1/2}}{T-P} + O(r) \;.
	\end{equation}
	In particular, $\tilde{q}$ does not tend to 0 as $r \rightarrow 0$.
\end{proposition}
\textbf{Proof}
Let $\overline{q}$ be the critical point of $\mu$ as in Lemma \ref{mufunction}. As $\mu(0)> (\epsilon_1+\epsilon_2)^{1/2}$ if $\epsilon$ is small enough, and $\mu$ is increasing in $[0, \overline{q}]$, then equation (\ref{muqtil}) has no solution in that interval. On the other hand, as $\mu$ is decreasing in $(\overline{q},q_{red}]$ with $\mu(q_{red})=0$, then (\ref{muqtil}) must have one root exactly in $(\overline{q},q_{red})$. 

In order to obtain (\ref{estqtil}) we rewrite (\ref{muqtil}) using definition (\ref{defmu}) along with (\ref{a13}) written in terms of variable $x$ defined by (\ref{defx}). Putting $\epsilon_1=\epsilon_2=0$ we obtain the approximate solution $x \approx q_{blue}$, which suggests us to define a new auxiliary variable $y$ as $ y = (x-q_{blue}) r^{-1/2}$. Substituting $x= q_{blue}+r^{1/2} y$ in (\ref{muqtil}) and making several simplifications, we get that (\ref{muqtil}) is equivalent to $H(r,y)=0$, where
\begin{eqnarray}
&&H(r,y) = G_1 r^{3/2} \cos^2 \theta (1-r^{1/2} y - q_{blue}) \nonumber \\&-& \frac{r^{1/2} \cos \theta [2R-S-T-r \cos \theta (T-P) - r \sin \theta (R-S)]}{q_{blue}+ r^{1/2}y} \nonumber\\
&+& r^{1/2} [(T-S) \cos \theta+(T-R) \sin \theta)] - (\cos \theta + \sin \theta)^{1/2} [1-r(\cos \theta+\sin \theta)] \nonumber\\
&-& (T-P)y \;.\nonumber
\end{eqnarray}
Repeating the argument with the implicit function theorem as in the proof of Proposition \ref{propestqred} we obtain (\ref{estqtil}).   $\blacksquare$

We now prove a monotonicity argument for $x_2(P_{121})-x_2(P_{131})$, but restricted to $(\tilde{q}, q_{red})$:
\begin{proposition} \label{monotdif}
	If $G_1 \geq 0$, $x_2(P_{121})-x_2(P_{131})$ is a decreasing function of $q$ in $(\tilde{q}, q_{red})$.
\end{proposition}
\textbf{Proof}
\[ \frac{d}{dq} x_2(P_{131}) \,=\, -\frac{ (q -\epsilon_2)(P-S) a'_{13}(q) + (R-a_{13}(q)) (P-S)}{[R-a_{13}(q)+(q -\epsilon_2)(P-S)]^2} \]
shows that $\frac{d}{dq} x_2(P_{131})$ is negative and $O(r)$ in $(\tilde{q}, q_{red})$. In order to conclude that, we are using the fact proved in Proposition \ref{propqtilde} that $\tilde{q}$ does not tend to 0 when $r \rightarrow 0$, which implies that both $R-a_{13}(q)$ and $a'_{13}(q)$ are $O(r)$ for $q \geq \tilde{q}$. The denominator is of course $O(1)$ due to the $(q -\epsilon_2)(P-S)$ term.

For $x_2(P_{121})$ we have
\[ \frac{d}{dq} x_2(P_{121}) \,=\, \frac{\epsilon_2(P-S) \mu'(q)}{[ \mu(q)+\epsilon_2(P-S)]^2} \,=\,  \frac {\mu'(q)}{\epsilon_2(P-S)} \left[ \frac{1}{1+\frac{\mu(q)}{\epsilon_2(P-S)}}\right]^2\;.\]
In $(\tilde{q},q_{red})$, $\mu'(q) =a'_{13}(q)-(T-P) <-\frac{1}{2}(T-P)$, where we are using again that $a'_{13}(q)$ is $O(r)$. Also, 
\begin{eqnarray}
\frac{1}{1+\frac{\mu(q)}{\epsilon_2(P-S)}} &>& \frac{1}{1+\frac{\mu(\tilde{q})}{\epsilon_2(P-S)}}\,=\, \frac{\sin \theta}{\sin \theta+ \frac{r^{-1/2} (\cos \theta+\sin\theta)^{1/2}}{P-S}}\nonumber \\
&>& \frac{\sin \theta}{r^{-1/2}[\sin \theta+ \frac{ (\cos \theta+\sin\theta)^{1/2}}{P-S}]} \geq K \sin \theta \, r^{1/2} \;, \nonumber
\end{eqnarray}
where $K= \max_{\theta \in [0, \pi/2]} \left(\sin \theta+\frac{(\cos \theta+ \sin \theta)^{1/2}}{P-S}\right)>0$.

Finally, we obtain, for $q\in(\tilde{q},q_{red})$,
\[\frac{d}{dq} x_2(P_{121}) < - \frac{1}{2}\, \frac{T-P}{P-S}\,K^2 \sin^2 \theta \;,\]
from which it turns out that $ \frac{d}{dq} (x_2(P_{121})-x_2(P_{131}))<0$.   $\blacksquare$

Putting together all known facts, we can now prove
\begin{theorem}[Black equilibrium, $G_1>0$]  \label{blacklarge}
	If $G_1>0$ and $q\in ((\epsilon_1+\epsilon_2)^{1/2},1]$, there is a single value $q_{black} \in (\tilde{q}, q_{red})$ such that the red, green and blue lines cross on the border of $B$. Moreover, the crossing is on $L_1$.
\end{theorem}
\textbf{Proof}
	By the properties of $\mu$ proved in Lemma \ref{mufunction}, and noticing that 
	\[x_2(P_{121}) = \frac{1}{1+ \frac{\epsilon_2 (P-S)}{\mu(q)}}\;,\]
	it is clear that the minimum of $x_2(P_{121})$ in $[0,\tilde{q}]$ is attained at one of the boundaries of the interval. But as $\mu(0)=O(1)$, and $\mu(\tilde{q})=(\epsilon_1+\epsilon_2)^{1/2}$, the minimum is attained at $\tilde{q}$ and its value is thus $1-O(r^{1/2})$.
	
	An easy calculation shows that the derivative of $x_2(P_{131})$ is negative in $[(\epsilon_1+\epsilon_2)^{1/2}, q_{red})$. It can be seen also that $x_2(P_{131})= 1-O(1)$ at $q= (\epsilon_1+\epsilon_2)^{1/2}$. Thus the maximum of $x_2(P_{131})$ is less than the minimum of $x_2(P_{121})$ in $ [(\epsilon_1+\epsilon_2)^{1/2}, \tilde{q}]$. This proves that the red and green lines do not cross on $L_1$ for $q \in [(\epsilon_1+\epsilon_2)^{1/2}, \tilde{q}]$.
	
	On the other hand, they do cross  for $q$ somewhere in $(\tilde{q},q_{red})$ because  we have already seen that at $q=q_{red}<q_{green}$ we have $x_2(P_{131})>0=x_2(P_{121})$. We have also just proved that the reverse holds at $q=\tilde{q}$. Uniqueness of this crossing in $(\tilde{q},q_{red})$ follows from Proposition \ref{monotdif}. Uniqueness in  $((\epsilon_1+\epsilon_2)^{1/2},1]$ is a consequence of Proposition \ref{posG1largeq}.   $\blacksquare$

Having settled the question of the black equilibrium for $G_1>0$, we remember that the question of whether the other equilibria are biological or not is already solved in Proposition \ref{P123} and Theorem \ref{biolgb}. The results of which equilibria are biological for $G_1>0$, all justified, are summarized in Table \ref{tabG1large}.

The elements for justifying the equilibria results of Table \ref{tabG10} for the case $G_1=0$ were already proved. It remains for us just the task organizing them. First of all, in the case $G_1=0$ we define $q_{black}$ to be equal to the common value $q_{blue}=q_{green}=q_{red}$. We then have
\begin{theorem}[Equilibria for $G_1=0$]
	If $G_1=0$, besides equilibria at the vertices of $B$ and the red equilibrium, which are always biological, these are the biological equilibria at each interval:
	\begin{itemize}
		\item The blue equilibrium is biological if and only $q \in [0, q_{black})$.
		\item If $q > (\epsilon_1+\epsilon_2)^{1/2}$, the green and black equilibria are biological if and only if $q \in (q_{black}, 1]$.
	\end{itemize}
\end{theorem}
\textbf{Proof}
	The assertions for the blue and green equilibria were already proved in Theorem \ref{biolgb}. In Theorem \ref{posG1largeq} we have already proved that the black equilibrium is biological for $q \in (q_{black}, 1]$. The only thing remaining to be proved is that the red, blue and green lines do not cross on the border of $B$ for $q \in ((\epsilon_1+\epsilon_2)^{1/2}, q_{black})$. 
	
	In fact, by Proposition \ref{ordsmallq} and the same argument in the proof of Theorem \ref{blacklarge}, we show that there is no crossing for $q \in ((\epsilon_1+\epsilon_2)^{1/2}, \tilde{q})$. No crossing for $q \in (\tilde{q},q_{red})$ is a consequence of Proposition \ref{monotdif}. Finally, for $(q_{red},1]$ the argument is Proposition \ref{posG1largeq}. So the red, green and blue lines only cross at the origin for $q=q_{black}$.   $\blacksquare$

The arguments necessary for proving validity of the equilibria results of Table \ref{tabG1small} are similar and simpler than the ones used for the other two cases, so that we will leave them to the reader.

\section{The dynamics}  \label{secdyn}
Replicator dynamics was studied from the point of view of the theory of Dynamical Systems in \cite{Zeeman}. Zeeman addressed mainly the robust cases, where robust means cases in which the dynamics remains unchanged for arbitrarily small changes of parameters. In particular, Zeeman showed that it was possible to classify the phase portraits of replicator dynamics by only knowing about the existence or not of equilibria at each face of the biological simplex, existence or not of an interior equilibrium, and the stability of each equilibrium. In the case of three strategies, the number of possible phase portraits was small enough so that each possibility could be shown. Zeeman's work was continued in \cite{Bomze83}, which included also the non-robust cases, thus obtaining a complete classification of all possible phase portraits for the replicator dynamics with three strategies. As we will also be interested in some non-robust cases, in this section we will refer to the classification by Bomze. In particular, we will see that, in most of the possible intervals and values of $G_1$ in Tables \ref{tabG1small} to \ref{tabG10}, our knowledge up to now can associate a single diagram in \cite{Bomze83} with compatible dynamics. In these cases the dynamics is then fully determined. In some other cases, indicated in the tables, there were three compatible diagrams. The enumeration of the diagrams in our tables is the same as in \cite{Bomze83}. The missing information, which led to doubt in determining the dynamics, is whether the black equilibrium is attractive, repulsive or neutral. Of course, for fixed values of parameters, it is an easy numerical task to linearize the dynamics around the black equilibrium, and by calculating eigenvalues of a $2 \times 2$ matrix, discover this information. But we do not have a general argument to show e.g. that the stability of the black equilibrium is the same for any choice of parameters in the last row of each of our tables.

Another information necessary for reading our Tables \ref{tabG1small} to \ref{tabG10} is that the letter R in front of the number of a diagram in \cite{Bomze83} means that one should take the corresponding diagram with all arrows reversed. In fact, the reader should notice that replacing matrix $A$ by $-A$ in equations (\ref{repleq}) has only the effect of reversing the orientation of all orbits.

In order to be able to reproduce the results about the diagrams in our tables, we must know about the stability of the equilibria on each side of $B$ whenever they are biological. On each side of $B$ one of the strategies is absent and we only need to study the one-dimensional replicator dynamics for two strategies. Results for this case are rather trivial, see e.g. \cite[p. 50]{Nowbook}, and only depend on the strategies being or not Nash equilibria. The results enumerated below are simple consequences of pairwise comparisons between elements of the pay-off matrix (\ref{irpdpayoff}).
\begin{enumerate}
	\item In the absence of strategy 3, strategies 1 and 2 are both strict Nash equilibria. Thus, for the replicator dynamics restricted to side $L_3$, the red equilibrium is unstable. Moreover, for any $q \in ((\epsilon_1+\epsilon_2)^{1/2},1]$ we can divide this side in four regions in which the fitness ranking is the following:
	\begin{itemize}
		\item Between $E_2$ and $P_{123}$, we have $f_2>f_1>f_3$.
		\item Between $P_{123}$ and $P_{233}$. we have $f_1>f_2>f_3$.
		\item Between $P_{233}$ and $P_{133}$. we have $f_1>f_3>f_2$.
		\item Between $P_{133}$ and $E_1$. we have $f_3>f_1>f_2$.
	\end{itemize}
	In particular, in a neighborhood of $P_{123}$, $f_3$ is the smallest fitness. This implies that $f_3< \phi$ in a neighborhood of $P_{123}$, with the consequence that orbits in the interior of $B$ close to $P_{123}$ must have $\dot{x}_3<0$, and flow towards $L_3$, which makes $P_{123}$ a saddle point. This property is particularly important, because in some cases it is what allows us to discard some diagrams in \cite{Bomze83} otherwise compatible.
	\item In the absence of strategy 1, we have two possibilities:
	\begin{itemize}
		\item If $q<q_{blue}$, the blue equilibrium is biological, and as strategies 2 and 3 are both strict Nash equilibria, then the blue equilibrium is unstable if the dynamics is restricted to the $L_1$ side. 
		\item If $q \geq q_{blue}$, the blue equilibrium is not biological, and only strategy 2 is a Nash equilibrium. Then all orbits on $L_1$ must flow into $E_2$.
		\end{itemize}
	\item In the absence of strategy 2, we also have two possibilities:
	\begin{itemize}
		\item If $(\epsilon_1+\epsilon_2)^{1/2}<q \leq q_{green}$, then the green equilibrium is not biological, and between strategies 1 and 3, only 3 is a Nash equilibrium. All orbits on $L_2$ must flow into $E_3$.
		\item If $q>q_{green}$, then the green equilibrium is biological. Because neither strategy 1, nor strategy 3 are Nash equilibria, then the green equilibrium is asymptotically stable when dynamics is restricted to $L_2$. 
	\end{itemize}
\end{enumerate}

An example can help clarify how we have obtained the results on the tables regarding the compatible diagrams and type of evolution of cooperation. We take as an example the first line in all three tables. We know that for $q>(\epsilon_1+\epsilon_2)^{1/2}$, but not too large, regardless of $G_1$ we will have as biological equilibria only the three vertices, and the red and the blue equilibria. By the above reasoning, $E_1$ must be a saddle point, whereas $E_2$ and $E_3$ are attractors, the red equilibrium is a saddle with outgoing orbits on $L_3$, and the blue equilibrium has outgoing orbits on $L_1$. Point $Q$ is not in the biological region.

\begin{figure*}
	\includegraphics[width=  \textwidth]{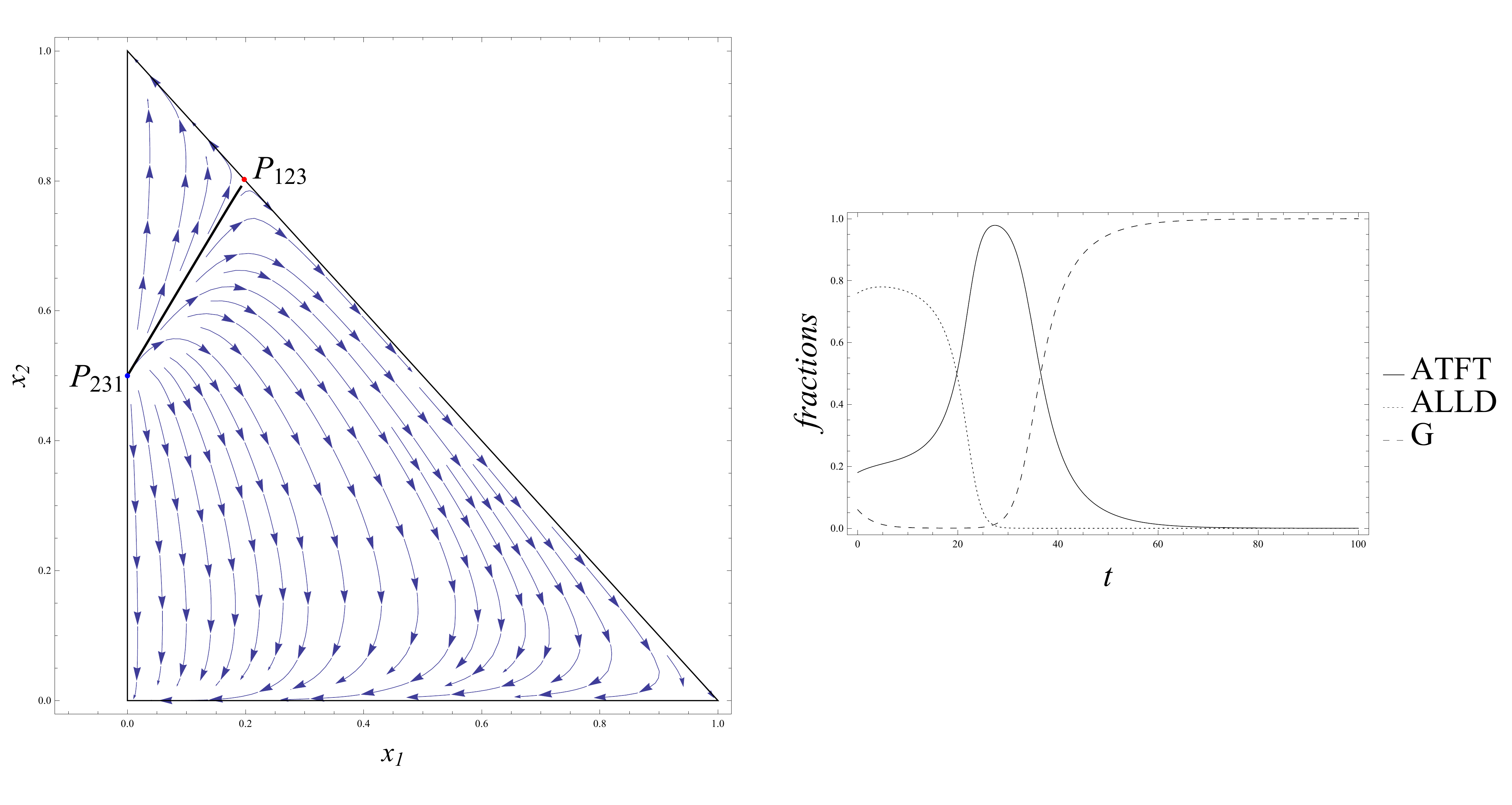}
	\caption{Phase portrait and graph of the fractions $x_1$, $x_2$ and $x_3$ as functions of time for the following parameter values: $T=5$, $R=4$, $P=2$, $S=0$, $\epsilon_1=.05$, $\epsilon_2=.10$, $q=.40$. The initial condition for the graphs is $(x_1,x_2)= (.18,.76)$. Observe the full evolution of cooperation.}
	\label{fullevo}       
\end{figure*}

Among the diagrams in \cite{Bomze83} not a single one is compatible with the above situation. But if we reverse the orbits, then diagrams 37 and 38 seem compatible. The only one which remains compatible when we take into account that interior orbits close to the red equilibrium must flow towards $L_3$ is 38R. In Figure \ref{fullevo} we show a plot of some numerically calculated orbits for a choice of parameters in one of the cases leading to diagram 38R. Notice that all orbits below the separatrix joining the blue and red equilibria lead to survival only of strategy G. This justifies why we have full evolution of cooperation. For orbits below the separatrix, but very close to the red equilibrium $P_{123}$ we can see the occurrence of the phenomenon in the experiment of \cite{NowSigNature}: an initial population with majority of ALLD, some ATFT and quite a few G individuals evolves to a population where only the G individuals are present, after passing through a transient in which the ATFT are almost the entire population. The phenomenon is illustrated by the graphs of fractions $x_1$, $x_2$ and $x_3$ as functions of time in Figure \ref{fullevo}.

As already mentioned, in the cases at the last line of each Table \ref{tabG1small} to \ref{tabG10} we could not find a rigorous argument for proving which of diagrams 12, 12R and 13 is the correct one. 

In the case $G_1>0$ (Table \ref{tabG1large}), we already know for  interval $(q_{blue},q_{green})$ that the only compatible diagram is 15R, in which the black equilibrium is unstable and, consequently, there is no evolution of cooperation. It is not reasonable that increasing $q$ will foster cooperation. In fact, larger values of $q$ will make the G individuals more susceptible to exploitation by ALLDs. Thus the natural conjecture is that if $G_1>0$ and  $q \in (q_{green},1]$ the black equilibrium will still be unstable and no evolution of cooperation will happen. If this conjecture is true, then the associated diagram must be 12R. The conjecture is supported also by numerical calculation of the eigenvalues of the linearized dynamics around the black equilibrium.

In the $G_1<0$ case we know that there is full evolution of cooperation until $q=q_{green}$ and only partial evolution for $q_{green}<q<q_{black}$, due to the green equilibrium destabilizing $E_3$. For larger values of $q$ the black equilibrium appears and we have no knowledge on its stability. Numerical calculation of eigenvalues suggests that in the case $G_1<0$ the black equilibrium is asymptotically stable and will attract all orbits in a region of positive area, which means weak evolution of cooperation and that the correct diagram should be 12.

Finally, in the $G_1=0$ case we may expect a situation intermediate between the other two cases. Numerical calculations suggest that the real part of the eigenvalues of the linearized dynamics around the black equilibrium may be 0. Numerically calculated orbits around the black equilibrium seem to be closed. The correct diagram would be 13, in which the interior equilibrium is a center, and evolution of cooperation would also be weak.

\section{Conclusions} \label{conclusions}
We have proved that the results of the computer experiment in \cite{NowSigNature} are true for a simplified version of that experiment in which, instead of 100 reactive strategies, we only have the three more prominent ones in the experiment: ATFT, ALLD and G. More precisely, if we define
\begin{equation}
\label{rigqtft}
q_{GTFT} \,=\, \min \{q_{green}, q_{blue}\}
\end{equation}
then for $q\in ((\epsilon_1+\epsilon_2)^{1/2},q_{GTFT})$ there exists a region $C \subset B$ with positive area such that the orbit of the replicator dynamics for any initial condition in $C$ will converge to $E_3$, i.e. only the G individuals will survive.

Equation (\ref{rigqtft}) should be thought as the rigorous version of the (\ref{molandergtft}). In fact, we say that (\ref{molandergtft}) is not precise because \cite{Molander}, neglecting $O(r)$ terms, defined GTFT as strategy $r(1,q)$ with $q$ \textit{close} to the value in (\ref{molandergtft}). Our equation (\ref{explqgreen}) shows in a precise sense that (\ref{molandergtft}) is indeed a good approximation. We also proved that (\ref{molandergtft}) is exact if and only if $G_1=0$, and that which between $q_{green}$ and $q_{blue}$ is the minimum depends on the sign of $G_1$.

We have also partially understood the population dynamics for values of $q$ larger than $q_{GTFT}$. We have seen that in some cases some weaker forms of cooperation evolution will still hold, but we have also seen that if $G_1>0$ and $q \in (q_{blue},q_{green})$ no evolution of cooperation is possible, because for almost all initial conditions only ALLD individuals will survive. The same situation probably holds also for $q \geq q_{green}$ and $G_1>0$, but probably it does not hold for $q>q_{black}$ and $G_1 \leq 0$.

%
%

\section*{Acknowledgements}
INR received a scholarship from Conselho Nacional de Desnevolvimento Cient\'{\i}fico e Tecnol\'ogico (CNPq, Brazil) during her master dissertation. AGMN was partially supported by Funda\c{c}\~ao de Amparo \`a Pesquisa de Minas Gerais (FAPEMIG, Brazil).

\bibliographystyle{plain}  
\bibliography{threestrat}   

\end{document}